\documentstyle[aps,epsfig,multicol]{revtex}

\begin{document}

\large


\newcommand{\om}{\omega}

\begin{center}

{\bf Quantum, Multi-Body Effects and Nuclear Reaction Rates in Plasmas}

V.~I.~Savchenko

{\it $^+$ Princeton University, PPPL, Forrestal Campus, Princeton, 
N. J. 08543}

\end{center}

\begin{abstract}
Detailed calculations of the contribution from off-shell
effects to the 
quasiclassical tunneling of fusing particles are provided. It is shown
that these effects   
change the Gamow rates of certain nuclear reactions in dense plasma by
several orders of magnitude.
\end{abstract}

\begin{multicols}{2}
\dimen100=\columnwidth \setlength{\columnwidth}{3.375in}

Methods of field theory do not straightforwardly apply to systems at
finite temperature.
Naive perturbation theory fails in the infrared region of energy at high
temperature but can be fixed by the partial HTL
resummation~\cite{htl-pisarski}. This approach was generalized
in~\cite{thermal-landsman}, and it was shown that fields without
dispersion relation~\cite{report-henning} are necessary for consistent
formulation of the theory.

The particle distribution over momenta, $f({\bf p})$, acquires
non-Maxwellian tail in 
thermodynamic equilibrium~\cite{tails-galitski} due to quantum
effects, while the 
particle distribution over energies, $\tilde{f}(\om)$, remains Maxwellian. An interesting
suggestion was made in Ref.~\cite{tails-cross-section-star}, that the
rates of various processes, including nuclear reactions, may be
significantly increased by the quantum tail of the {\em momentum}
distribution. Non-linear integral equation for the rate of inelastic
processes as well as a final formula for the ionization rate 
was obtained in \cite{tails-cross-section-star}.
The formula for the nuclear reaction rates was found
in \cite{neutrino-vlad} under assumption, that it is correct to
substitute the particle momentum in the form 
$\epsilon_{\bf p}=p^2/2m$ into the known quasiclassical cross-section
$\sigma(\om)$ and then average it over the particle distribution
over momenta $f({\bf p})$ rather than the distribution over energies
$\tilde{f}(\om)$. This
procedure becomes even more unclear
, if we take into account the fact, that the
momentum, ${\bf p}$, and energy, $\om$, of a {\em colliding}
particle are {\em independent} variables 
not connected by the usual dispersion relation $\delta(\om -
{\bf p}^2/2m)$~\cite{tails-galitski,statistical-mechanics-kadanov,tails-cross-section-star}.

Note, that the averaging of $\sigma(\om)$ over the distribution
over energies $\tilde{f}(\om)$, which is Maxwellian, would
give the usual reaction rates~\cite{rate-gamow} rather than the rates found in
\cite{neutrino-vlad}, which are accelerated by many orders of
magnitude in certain regimes. 
One can
argue~\cite{neutrino-vlad} 
that averaging over momentum may not be completely wrong, because the
dependence of the 
scattering amplitude on energy can be neglected in the
gaseous
approximation~\cite{statistical-abrikosov,trydi-galitski}. However,
the connection between the scattering 
amplitude and the rate of nuclear reactions is not clear.
Neither is it obvious that this
argument can straightforwardly lead to correct answer in the case of
essentially non-linear processes of nuclear reactions. 
It was shown that
non-linearities can be very subtle, subject to various non-trivial
cancellations~\cite{rates-brown-sawyer}. 
It is therefore important to find the
nuclear reaction rates from the first principles.

In this paper we rigorously
solve the quasiclassical problem of tunneling through the potential
barrier. Our result demonstrates
that the cancellations of the type found in~\cite{rates-brown-sawyer}
do not play a role
in the case under consideration. We discuss this question in
more details in the future publication~\cite{two-rates-vlad}.

The tunneling particles undergo simultaneous collisions
with other particles of the plasma maintained in thermodynamic
equilibrium. The plasma 
is assumed to be fully ionized. We use the Green--function 
technique, introduced by Keldysh and
Korenman~\cite{landau-10,diagramms-keldish}, and do
not rely on any {\em ad hoc} assumptions~\cite{neutrino-vlad} about
any averaging procedure. 
Our final result for the nuclear reaction rate~\cite{rates-vlad}
agrees with that postulated in \cite{neutrino-vlad}, apart from the
coefficient depending on the masses of colliding and reacting particles.

A number of different off-shell processes in plasmas were investigated
within a framework of thermal field theory.
Emission rate of soft photons from hot plasmas was calculated
in~\cite{soft-photons-henning}. Using
full spectral functions allowed to eliminate unphysical divergences
and extend the range of validity of the results obtained by
HTL resummation by taking into account off-shell particle propagation.
Landau damping was shown to lead to anomalous relaxation of the
particle distribution
function~\cite{quasi-blaizot-iancu}.
Off-shell effects due to particle collisions were treated within Born
approximation in~\cite{nonequil-danielewicz}. The relaxation rate,
calculated 
with the help of full spectral 
functions, is slower approximately by a factor of 2 as
compared to the rates calculated by solving the Boltzman
equation~\cite{nonequil-danielewicz}. 
This relaxation rate is, however, an integral quantity, 
insensitive to the details of 
changes of the kinetic Green function due to inclusion of off-shell
effects. 

In contrast to the relaxation rate, inherent non-linearity of the tunneling
problem makes the average nuclear reaction rate very sensitive to the
behaviour of this function at large momenta. This explaines why we
should expect off-shell effects to lead to
significant corrections to the nuclear reaction 
rates. Indeed, we find that certain reactions are accelerated by many
orders of magnitude as compared to the 
calculations, performed using spectral functions with zero
width~\cite{rates-brown-sawyer}. We solve the tunneling problem,
neglecting Salpeter corrections due 
to screening, and assume that, the nuclear
transformations are unaffected by the medium.

Kinetic properties of the particles are described by the Green function
$G^{-+}(X_1,X_2)$, where $X=(t,{\bf
x})$~\cite{diagramms-keldish}. Equations
for $G^{-+}$ include the
interaction between the particles as well as their interaction with
the external field. Therefore, the rate of change of this
function, evaluated at any point after the barrier, will give the
tunneling rate $K({\bf x}, t)$ for the particles colliding
simultaneously with other particles~\cite{rates-vlad}:
\begin{eqnarray}
K({\bf x}, t)=\left[(\partial_{t_1} - \partial_{t_2}) G^{-+}(t_1 {\bf
x_1}; t_2 {\bf x_2})\right]_{X_1 \rightarrow
X_2=X}&& \label{rate-general}
\end{eqnarray}

Now we will proceed with solution of the kinetic equations for
$G^{-+}(X_1,X_2)$. They can be written as Dyson equations for
$G^{\alpha \beta}_{12}$~\cite{landau-10}:
\begin{equation}
G^{\alpha \beta}_{12}=G^{(0)\alpha \beta}_{12} + \int G^{(0)\alpha
\gamma}_{14} \Sigma^{\gamma \delta}_{43} G^{\delta \beta}_{32} d^4 X_4
d^4 X_3, \label{kinetic-general}
\end{equation}
where we use subscripts $1..4$ to denote dependence on $X_1..X_4$. 
Dependence on $G^{(0)\alpha \beta}_{12}$ can be
eliminated~\cite{landau-10} by acting 
on both sides of the Eq.~(\ref{kinetic-general}) with the operator
\begin{eqnarray*}
&&\hat{G}^{-1}_1=i\frac{\partial}{\partial t_1} + \frac{\Delta_1}{2 m} -
U(z_1)\equiv i\frac{\partial}{\partial t_1} + \frac{\Delta_1^{\perp}}{2 m} +
\hat{L}(z_1), 
\end{eqnarray*}
where $\Delta_1^{\perp}\equiv \partial^2_{x_1} + \partial^2_{y_1}$.
Since we are interested in the steady state solution,
one can see that, the properties of this operator can be fully
accounted for if we 
introduce the function $g(z,k)$, such that
\begin{equation}
\left[\frac{1}{2m} \frac{\partial^2}{\partial z^2} -
U(z) \right] 
g(z,k)=-\epsilon_k g(z,k),
\end{equation}
where $\epsilon_k=k^2/2m$.
Since the operator $\hat{L}$ is Hermitian, we will use the
property of its eigen-functions $g(z,k)$:
\begin{equation}
\int_{-\infty}^{\infty} g(z,k_1) g^*(z,k_2) dz= 2 \pi \delta(k_1 - k_2)
\label{orthogonality} 
\end{equation}

Now let us solve for $G^{(0)-+}_{12}$, which satisfies
\begin{equation}
\hat{G}^{-1}_1 G^{(0)-+}_{12}=0. \label{kinetic-zero}
\end{equation}
It is easier to find $G^{(0)-+}_{12}$ by using its definition
\begin{equation}
G^{(0)-+}_{12}=i\left<\hat{\psi}_2^{\dagger} \hat{\psi}_1 \right>,
\label{mp-definition} 
\end{equation}
where $\hat{\psi}^{\dagger}, \hat{\psi}$ are the Heisenberg operators of
creation and annihilation of the particles at a point $X$, and $<..>$
means quantum and statistical averaging. 
Since $\hat{\psi}$ evolves according to $\hat{G}^{-1} \hat{\psi}=0$,
one can see, that it is equal to
\begin{equation}
\hat{\psi}(X)=\sum_{{\bf q}_{\perp} k} \hat{a}_{{\bf q}_{\perp} k}
g(z,k) e^{-i 
\epsilon_{{\bf q}_{\perp} k} t + i {\bf q}_{\perp}\cdot {\bf
x}^{\perp}} \label{psi}
\end{equation}
Here $\hat{a}_{{\bf q}_{\perp} k}$ is the annihilation operator of the
particle with momentum $({\bf q}_{\perp},k)$, and
$\epsilon_{{\bf q}_{\perp} k}\equiv {\bf q}_{\perp}^2/2m + k^2/2m$.

We substitute $\psi(X)$ from (\ref{psi}) into
Eq.~(\ref{mp-definition}) and obtain
\begin{eqnarray*}
G^{(0)-+}_{\om {\bf q}_{\perp}}(z_1,z_2)=2 \pi i
\int_{-\infty}^{\infty} \frac{dk}{2 \pi}&& \hspace*{15mm}
\end{eqnarray*}
\begin{equation}
 \delta(\om - \epsilon_{{\bf q}_{\perp} k} + \mu) n({\bf q}_{\perp} k)
g(z_1,k) g^*(z_2,k), \label{mp-zero}
\end{equation}
where we denote $\left<a^{\dagger}_{{\bf q}_{\perp} k} a_{{\bf
q}_{\perp} k}\right>\equiv n({\bf q}_{\perp} k)$; $\mu$ is the
chemical potential.

It is easy to see that, in thermodynamic equilibrium we are
considering, the function $n({\bf q}_{\perp} k)$ is equal to Fermi
distribution $n_F(\epsilon_{{\bf q}_{\perp} k})$. 
To convince ourselves we take
$G^{(0)-+}_{\om {\bf q}_{\perp}}(z_1,z_2)$ at such values of
$z_1,z_2$, when $U(z)$ is negligible and $g(z,k)\Rightarrow
exp(ikz)$. One more Fourier transform $\int
exp[-iq(z_1 - z_2)] d(z_1 - z_2)$ will give $G^{(0)-+}_{\om {\bf q}}$,
which has to coincide with the known expression~\cite{landau-10}
$G^{(0)-+}_{\om {\bf q}}=2 \pi i \delta(\om - \epsilon_{\bf
q} + \mu) n_F(\epsilon_{\bf q})$.

Before proceeding further with solving for $G^{-+}(X_1,X_2)$ we first
find $K^{(0)}$, which is the tunneling rate obtained by using
$G^{(0)-+}_{12}$ in Eq.~(\ref{rate-general}). 
Since we are interested in $K^{(0)}$ behind the barrier, where $U(z)=0$, we
can exchange $\partial_{t_1}$ for $i\Delta_1/2m$ as is clear from the
form of the operator $\hat{G}^{-1}_1$ and Eq.~(\ref{kinetic-zero}).
Then the formula for the rate
$K^{(0)}$ takes the form:
\begin{equation}
K^{(0)}(z)=\frac{i}{2m}\int \frac{d\om}{2 \pi} \int\frac{d^2{\bf
q_{\perp}}}{(2 \pi)^2} \left[\frac{\partial^2}{\partial \zeta^2}
G^{(0)-+}_{\om {\bf q}_{\perp}}(z,\zeta)\right]_{\zeta=0}, \label{rate-zero}
\end{equation}
where $z=(z_1 + z_2)/2$, $\zeta=z_1 - z_2$.

In deriving (\ref{rate-zero}) we expressed $G^{(0)-+}_{12}$ through
its Fourier transform $G^{(0)-+}_{\om {\bf q}_{\perp}}(z,\zeta)$ with
respect to ``fast'' space ${\bf r}_{\perp}={\bf r}^{\perp}_1 - {\bf
r}^{\perp}_2$ and time $\tau=t_1 - t_2$
variables and took the limit ${\bf r},\tau \rightarrow 0$ as
prescribed by Eq.~(\ref{rate-general}).

Finally we need to know the function $g(z,k)$ in the region behind the
barrier. We will use the standard quasiclassical
expression~\cite{landau-3}
\begin{equation}
g(z,k)=\frac{C_k}{\tilde{q}(z,k)^{1/2}} exp\left[i\int_{z_*}^z
\tilde{q}(z',k) dz' \right] \label{g-quasicl} 
\end{equation}
\begin{equation}
C_k=A_k exp\left[-\int_{z_a}^{z_b(k)}
\tilde{q}(z',k) dz'\right]  \label{ck}
\end{equation}
\begin{equation}
\tilde{q}(z,k)=\sqrt{2m(\epsilon_k - U(z))} \label{qk}
\end{equation}
The square of the exponential factor in Eq.~(\ref{ck}) is the
tunneling coefficient, $\tilde{W}(k)$~\cite{landau-3}.
We multiply it by a factor $A_k=(S(\epsilon_k)/\epsilon_k)^{1/2}$, which
takes into account nuclear 
physics effects, not considered here. $S(\epsilon_k)$ is the astrophysical
factor~\cite{fusion-adelberger}. 

To find $K^{(0)}$ from Eq.~(\ref{rate-zero}) we substitute
Eqs.~(\ref{g-quasicl})-(\ref{qk}) in Eq.~(\ref{mp-zero}), use
$z_1=z+\zeta/2$, 
$z_2=z-\zeta/2$ and take
$\partial_{\zeta}^2$ derivative in the quasiclassical sense keeping
only zero-th order terms. The result of the differentiation is
\begin{equation}
\left[\frac{\partial^2}{\partial \zeta^2} \left[g(z_1,k) g^*(z_2,k)\right] \right]_{\zeta=0}=-\tilde{q}(z,k) C_k^2
\label{derivative} 
\end{equation}
Note, that $\tilde{q}(z,k)=k$ is
true behind the barrier. 

The problem of tunneling collision between particles $m_1$ and $m_2$
can be reduced to the one-dimensional motion of a particle with
reduced mass
$m_r=m_1 m_2/(m_1 + m_2)$ in the external
potentianl $U(r)$~\cite{landau-3}. We make this reduction in the original
Hamiltonian and then second
quantize it. 

We need to perform calculations of $G^{(0)-+}$ and $K^0$ similar to
those outlined 
above. However, now we have to do 3-D calculations, using spherical
coordinates. So we replace $\hat{L}(z)$-operator with $\hat{L}_{\bf
x}=\Delta_{\bf x}/2m_r$, denote its radial part in spherical
coordinates by $\hat{L}_r$ and
introduce the function $g({\bf x})$ according to 
\begin{eqnarray}
&& g({\bf x})=\int\frac{k^2 dk}{(2\pi)^3} g_k({\bf x})\\
&& g_k({\bf x})=\sum\limits_{lm} g_{kl}(r)
Y_{lm}(\theta \phi) \\
&&\left(\hat{L}_r - U_l(r)\right) g_{kl}(r)=-\epsilon_{\bf k}
g_{kl}(r),
\label{gk-definition}
\end{eqnarray}
where $Y_{lm}(\theta \phi)$ is a spherical harmonic and
$U_l(r)=U(r) + \frac{l(l+1)}{2m_r r^2}$.
Quite analogously to Eq.~(\ref{psi}) we express $\psi$ through $g({\bf
x})$ as
\begin{equation}
\psi({\bf x}_1, t_1)=\sum\limits_{k_1 l_1 m_1} \hat{a}_{k_1 l_1}
g_{k_1 l_1}(r_1) Y_{l_1 m_1}(\theta_1 \phi_1) 
e^{-i \epsilon_{{\bf k}_1} t_1}.
\label{psi-spherical}
\end{equation}

Now using Eqs.~(\ref{mp-definition}), 
(\ref{gk-definition}), (\ref{psi-spherical}) and taking into account
the contribution only from spherically symmetric modes, we obtain $G^{(0)-+}$  
\begin{eqnarray*}
G^{(0)-+}_{\om}({\bf x}_1,{\bf x}_2)=2 \pi i
\int_{0}^{\infty} \frac{k^2 dk}{(2 \pi)^3}&& \hspace*{15mm}
\end{eqnarray*}
\begin{equation}
 \delta(\om - \epsilon_{\bf k} + \mu) n_F(\epsilon_{\bf k})
g_k({\bf x}_1) g_k^*({\bf x}_2). \label{mp-zero-spherical}
\end{equation}

Rewrite the rate $K^0$, Eq.~(\ref{rate-zero}), as
\begin{equation}
K^{(0)}({\bf x})=\frac{i}{2m_r}\int \frac{d\om}{2 \pi} 
\left[\Delta_{{\bf x}_1} G^{(0)-+}_{\om}({\bf x}_1,{\bf
x}_2)\right]_{{\bf x}_1 \rightarrow {\bf x}_2}, 
\label{rate-zero-spherical}
\end{equation}
In the WKB approximation we get for the function $g_k(r)$ the
following formula
\begin{equation}
g_k(r)=\frac{1}{r} \chi(r)
\end{equation}
\begin{equation}
\chi(r)=\frac{\bar{C}_k}{\tilde{q}(r,k)^{1/2}} exp\left[i\int_{r_0}^r
\tilde{q}(r',k) dr' \right] \label{chi-quasicl} 
\end{equation}
\begin{equation}
W(k)=exp\left[-\int_{r_0}^{\alpha/\bar{\epsilon}_{\bf k}}
\tilde{q}(r',k) dr'\right]  \label{ck-spherical}
\end{equation}
\begin{equation}
\tilde{q}(r,k)=\sqrt{2m_r(\bar{\epsilon}_k - U(r))}, \label{qk-spherical}
\end{equation}
where the bar denotes dependence on the reduced mass,
$\bar{\epsilon}_k=k^2/2m_r$, $\bar{C}_k^2=\bar{A}_k^2 W(k)$,
$\alpha\equiv Z_1 Z_2 e_1 e_2$ and
$\bar{A}_k^2=r_0^2 S(\bar{\epsilon}_k)/\bar{\epsilon}_{\bf k}$. 
We evaluate the derivative of Eq.~(\ref{rate-zero-spherical})
in the quasiclassical sense
\begin{equation}
\left[g_k^*({\bf x}_2)\Delta_{{\bf x}_1} g_k({\bf x}_1)\right]_{{\bf
x}_1\rightarrow {\bf x}_2=r_0}=-k \frac{\bar{C}_k^2}{r_0^2} 
\label{gk-derivative}
\end{equation}
and obtain the final result for $K^{(0)}$ 
\begin{equation}
K^{(0)}=n_0 \int_{0}^{\infty}
\frac{4 \pi k^2 dk}{(2 \pi)^3} \bar{n}_M(k) \frac{k}{m_r}
\frac{S(\bar{\epsilon}_k)}{\bar{\epsilon}_k} W(k), \label{rate-zero-final}
\end{equation}
where $\bar{n}_M(k)$ is the Maxwell distribution which depends on $m_r$.

It agrees with the 
answer obtained in~\cite{rate-gamow} by averaging the quasiclassical
tunneling factor $W(k)$ over the Maxwell distribution,
$\left<n_M(v)\sigma(m_rv^2/2)v\right>$. Hence, we see, that the
dispersion relation $\delta(\om - \epsilon_{\bf q})$ was implicitly
used in \cite{rate-gamow}, which corresponds to an approximation of
instantaneous, two-body collisions. Now we will not make this 
assumption and proceed with finding the 
tunneling rate, $K$, of the particles, which always collide and hence are
never ``in between collisions''. 

By using~\cite{landau-10,statistical-mechanics-kadanov}, we find 
\begin{eqnarray*}
\left(\frac{q}{m}\frac{\partial}{\partial z}  +  U'(z)
\frac{\partial}{\partial q}\right) G^{-+}_{\om {\bf
q}}(z)=- \gamma_{\om {\bf q}}(z) G^{-+}_{\om {\bf q}}(z)
\end{eqnarray*}
\begin{equation}
-\gamma_{\om {\bf
q}}(z) n_F(\om) \left(G^R_{\om {\bf q}}(z) - G^A_{\om {\bf
q}}(z)\right), \label{kinetic-inter}
\end{equation}
where we denote $\Sigma^{R} - \Sigma^{A}\equiv i \gamma_{\om
{\bf q}}(z)\equiv i \gamma_{\om {\bf q}_{\perp}}(q,z)$.
Note, that $\gamma_{\om {\bf q}}(z)$ is a non-linear function of
$G^{\alpha \beta}$~\cite{statistical-mechanics-kadanov,landau-10}.    
We now analyze Eq.~(\ref{kinetic-inter}) qualitatively,
which will help us to find that part of the
solution, which makes the largest contribution to $K$.

In the region {\em away} from the barrier we can neglect the
LHS of Eq.~(\ref{kinetic-inter}), and obtain:
\begin{equation}
\tilde{G}^{-+}_{\om {\bf q}}(z)=-n_F(\om)\left(G^R_{\om {\bf q}}(z) -
G^A_{\om {\bf q}}(z)\right) \label{mp-tilde}
\end{equation} 
Therefore, Eq.~(\ref{kinetic-inter}) will
allow us to propagate this solution into the region {\em behind} the
barrier.

We can find the first integral and formally integrate
Eq.~(\ref{kinetic-inter}) along trajectories, 
which will lead to a sum of solutions of homogeneous and
"inhomogeneous" equations (due to the last term in
(\ref{kinetic-inter})). As can be seen from Eq.~(\ref{kinetic-inter}),
"inhomogeneous" solution will involve the integral of the product of
$\gamma_{\om {\bf q}}(z)$ and $G^R_{\om {\bf q}}(z)$ or $G^{R*}=G^A$. 
Qualitatively,
it means, that in the region of $z$ behind the barrier it will be
proportional to a product of at least two exponential 
factors $W(k)$. This is so, because both $\gamma_{\om
{\bf q}}(z)$ and $G^R_{\om {\bf q}}(z)$ will depend on the integral of
the product $g(z_1,k) g^*(z_2,k)$ and such product is $\propto W(k)$
as can be seen from Eqs.~(\ref{g-quasicl})-(\ref{qk}).

Explicitely, the expression for the retarded Green function can be
obtained through the usual resummation~\cite{statistical-abrikosov}
\begin{equation}
G^R_{\om {\bf q}}(z)=\frac{1}{\om - \epsilon_{\bf q} -
\Sigma^R_{\om {\bf q}}(z)}
\label{gr}
\end{equation}
Since the advanced function, $G^A$, is related to $G^R$ by
$G^A={G^{R*}}$, we find that their difference is
\begin{eqnarray}
G^R-G^A&=&2Im\left(\frac{1}{\left(G_0^R\right)^{-1} - \Sigma^R}\right)
\nonumber\\ 
&=&\frac{\gamma}{\left(G_0^R\right)^{-2} + \left(Im \Sigma^R\right)^2} 
\label{gr-differ}
\end{eqnarray}

Now we will show that $\gamma$ is $\propto W(k)$, which will allow us
to neglect  the ``inhomogeneous'' term in
Eq.~(\ref{kinetic-inter}), as being second order in $W(k)$ since
$G^R-G^A$ is also 
proportional to $W(k)$. 
First, note that Eq.~(15) is written in the Furry
representation. The Furry representation is needed,
since we are interested in the behaviour of all the
functions in the external field $U(z)$. This means that
the diagrammatic expansion of the operator
$\gamma=2Im\Sigma^R$ in Eq.~(\ref{kinetic-inter}) is built upon zero-th order
functions, which incorporate dynamics in the external potential and
hence are exponentially suppressed in the region behind the barrier. This
is in contrast to the usual expansion which utilizes zero-th order
functions of free particles. (Such representation is usually called
Furry representation in the literature.) In other words, these
functions resolve the operator
\begin{equation}
\left(\frac{q}{m}\frac{\partial}{\partial z}  +  U'(z)
\frac{\partial}{\partial q}\right) \label{operator-z}
\end{equation}
which is the left hand side of Eq.~(\ref{kinetic-inter}). Therefore,
they are exponentially suppressed in the region of $z$ behind the
barrier. This can be also seen from the explicit expressions for the
Green functions. The function $G^{(0)-+}$, is
written in Eq.~(\ref{mp-zero}), and is proportional to the product
$g(z_1, k) g^*(z_2, k)$. As discussed above, this product is
exponentially suppressed behind the barrier. Analogously, the function
$G^{(0)+-}$ can be obtained from 
$G^{(0)-+}$ by substitution $n({\bf q}_{\perp} k) \rightarrow (1 -
n({\bf q}_{\perp} k))$, and is also suppressed exponentially, since
its dependence on $z$ is the same.
This is an important property, which we will make use of, while
analyzing the behaviour of $\gamma$ in the region behind the barrier.

Now we can write explicitely
expression of the first Born diagram for $\gamma_{\om {\bf
q}_{\perp}}(z_1, z_2)$ in the Furry representation from the Fig.~1 b)
of Appendix A. There is a slight difference from the case of
free zero-th order functions. Namely, there are additional integrals
over intermediate $z_3, z_4$-points in the loop (see Appendix A for notations):

\end{multicols}

\begin{eqnarray}
-i\Sigma^{-+}_{\om {\bf p}_{\perp}}(z_1, z_2)=\int G^{-+}_a&&(\om, {\bf p}_{\perp}-{\bf q}_{\perp};
z_1, z_2) G^{+-}_b(\om, {\bf p}_1^{\perp}; z_4, z_3) G^{-+}_b(\om,
{\bf p}_1^{\perp} - {\bf q}^{\perp}; z_3, z_4) \nonumber\\ 
&&V({\bf q}_{\perp}; z_1 - z_4)  V({\bf q}_{\perp}; z_2 - z_3) \frac{d\om_{{\bf p}_1} d{\bf p}_1^{\perp}}{(2
\pi)^3} \frac{d\om_{{\bf q}} d {\bf q}_{\perp}}{(2
\pi)^3} dz_3 dz_4, \label{sigmafurry-mp}
\end{eqnarray}

\begin{multicols}{2}
\dimen100=\columnwidth \setlength{\columnwidth}{3.375in}

Since all the Green functions in Eq.~(\ref{sigmafurry-mp}) are
exponentially suppressed behind the barrier, we will obtain the largest
contribution, if we restrict integration over $z_3$ and $z_4$ to the
region away from the barrier, $z>z_*$. The functions $G^{+-}_b,
G^{-+}_b$ are not exponentially suppressed there.
However, note that $\Sigma^{-+}_{\om {\bf
p}_{\perp}}(z_1, z_2)$ depends on $z_1, z_2$ through the function
$G^{-+}_a$ in Eq.~(\ref{sigmafurry-mp}).
One can see that behaviour of $\Sigma^{-+}_{\om {\bf q}_{\perp}}(q, z)$ in
the region of $z$ behind the barrier will be determined by the
behaviour of $G^{-+}_a(\om, {\bf p}_{\perp}-{\bf q}_{\perp};
z_1, z_2)$, Wigner-transformed as in Eq.~(\ref{gamma-qz}). The same
argument applies to $\Sigma^{+-}$, which will be determined by
$G^{+-}_a$. Therefore, $\gamma=\Sigma^{-+} - \Sigma^{+-}$, which is 
\end{multicols}
\begin{equation}
\gamma_{\om {\bf q}_{\perp}}(q, z)=\int exp[- i q (z_1 - z_2)] \left(
\Sigma^{-+}_{\om {\bf q}_{\perp}}(z_1, z_2) - \Sigma^{+-}_{\om {\bf q}_{\perp}}(z_1, z_2)\right)d(z_1 - z_2) \label{gamma-qz}
\end{equation}
\begin{multicols}{2}
\dimen100=\columnwidth \setlength{\columnwidth}{3.375in}
will depend on the behaviour of these Green functions behind the barrier.

But we know
that, since both $G^{(0)-+}_a$ and $G^{(0)+-}_a$ resolve the operator
(\ref{operator-z}), they are
exponentially suppressed behind the barrier. (This is also proven by
their explicit expressions, as explained above).
Therefore, $\gamma_{\om
{\bf q}_{\perp}}(q, z)$ is also exponentially suppressed there.

More formally, we can obtain explicit expression for the $\gamma$ as
far as its dependence on $z$ is concerned. First, we have to simplify
Eq.~(\ref{sigmafurry-mp}); then apply similar arguments to $\Sigma^{+-}$,
which will allow to find $\gamma$. 

We are interested in the region of $z\equiv (z_1 + z_2)/2$
behind the barrier.  We can make this
region rather small, compared to the region away from the
barrier, by choosing potential $U(z)$ to be infinite for
$z<z_0$. Here $z_0$ is an arbitrary, finite point behind the barrier. 
This procedure does not restrict generality of this argument, since
the actual region behind the Coulomb barrier of a fusing particle
is much smaller than the region away from the barrier.
Then, we can approximate $V({\bf q}_{\perp}, z_1 - z_4)$ by $V({\bf q}_{\perp},
z - z_4)$, as well as $V({\bf q}_{\perp}, z_2 - z_3)$ by
$V({\bf q}_{\perp}, z - z_3)$ in Eq.~(\ref{sigmafurry-mp}). As we explained
above, the main 
contribution to the 
integral over $z_3, z_4$ comes from the region of $z>z_*$ (the region
away from the barrier). In this
region, the functions $g_b(z_3, k), g_b(z_4, k)$, entering
$G^{(0)+-}_b, G^{(0)-+}_b$ in Eq.~(\ref{sigmafurry-mp}), can be well
approximated by $exp(ikz_3), exp(ikz_4)$. After this approximation,
the region of integration can be extended from $-\infty$ to $\infty$,
which will 
introduce only an error in the numerical coefficient on the order of unity.
By using the explicit forms of 
$G^{(0)-+}, G^{(0)+-}$ from Eq.~(\ref{mp-zero}), we can carry
explicitely integrations over frequencies. Then we find

\end{multicols}

\begin{eqnarray}
-i\Sigma^{-+}_{\om {\bf p}_{\perp}}(z_1, z_2)=&&-4\pi i\int\int
\frac{d{\bf p}_1}{(2 \pi)^3} \frac{d{\bf q}}{(2 \pi)^3}
\int\limits_{-\infty}^{\infty} \frac{dk}{(2 \pi)} n^a_{{\bf p}_{\perp}
- {\bf q}_{\perp} k} n^b_{{\bf p}_1}   
g_a(z_1, k) g_a^*(z_2, k)  \nonumber \\
&&V_{\bf q}^2 \delta(\om_p - \epsilon_{{\bf p}_1}^b + \epsilon_{{\bf
p}_1 - {\bf q}}^b -   
\epsilon_{{\bf p}_1^{\perp} - {\bf q}^{\perp}, k}^a) \label{sigmamp-fin}
\end{eqnarray}  

In a similar manner, we find the answer for the $\Sigma^{+-}$:
\begin{eqnarray}
-i\Sigma^{+-}_{\om {\bf p}_{\perp}}(z_1, z_2)=&&-4\pi i\int\int
\frac{d{\bf p}_1}{(2 \pi)^3} \frac{d{\bf q}}{(2 \pi)^3}
\int\limits_{-\infty}^{\infty} \frac{dk}{(2 \pi)} 
(1-n^a_{{\bf p}_{\perp} - {\bf q}_{\perp} k} - n^b_{{\bf p}_1})
n^b_{{\bf p}_1 - {\bf q}}   
g_a(z_1, k) g_a^*(z_2, k)  \nonumber \\
&&V_{\bf q}^2 \delta(\om_p - \epsilon_{{\bf p}_1}^b + \epsilon_{{\bf
p}_1 - {\bf q}}^b -   
\epsilon_{{\bf p}_1^{\perp} - {\bf q}^{\perp}, k}^a) \label{sigmapm-fin}
\end{eqnarray} 

\begin{multicols}{2}
\dimen100=\columnwidth \setlength{\columnwidth}{3.375in}

The expression for $\gamma$ follows from Eqs.~(\ref{gamma-qz},
\ref{sigmamp-fin}, \ref{sigmapm-fin}). We see that its dependence on
$z$ is determined by functions $g(z,k)$, which are exponentially
suppressed behind the barrier. This completes the proof and allows us
to neglect the ``inhomogeneous'' term in the Eq.~(\ref{kinetic-inter}).

The solution of the "homogeneous" equation,
$G^{-+}_f$, involves only one factor of $W(k)$, since $G^{-+}_f
\propto g g^*$, see Eq.~(\ref{kinetic-inter}). Therefore, to obtain
$K$ we will find
only the solution to the "homogeneous" equation.
We supplement this homogeneous equation with the boundary
condition, $\tilde{G}^{-+}$ from Eq.~(\ref{mp-tilde}) imposed at
$z=z_*$ away from the barrier. 

Since the largest contribution to $K$ is made by $G^{-+}_f$ one can
see, that we can use Eq.~(\ref{rate-zero}) to find $K$ if we substitute
$G^{-+}_f$ instead of $G^{(0)-+}$ in (\ref{rate-zero}).
Therefore, we find $G^{-+}_f$ with
dependence on $z_1, z_2$ from the very begining by acting on
Eq.~(\ref{kinetic-general}) with the operator
\begin{equation}
\int \int g(z_1,k_1) g^*(z_2,k_2) \frac{d k_1}{2 \pi} \frac{d k_2}{2
\pi} \label{transformation}
\end{equation}
We use the boundary condition,
Eq.~(\ref{mp-tilde}) and obtain the
following answer for $G^{-+}_f$: 
\begin{equation}
i n_F(\om)\int g_k(z_1)\delta_{\gamma}(\om -
\epsilon_{{\bf q}_{\perp}k}) g^*_k(z_2) dk \label{mp-final}
\end{equation}
where $g_k(z)\equiv g(z,k)$ and we used for $\delta_{\gamma}$~\cite{statistical-mechanics-kadanov} 
\begin{eqnarray}
2 Im G^R_{\om {\bf q}_{\perp}}(k)=\frac{\gamma_{\om {\bf q}_{\perp}}(k)}{(\om -
\epsilon_{{\bf q}_{\perp}k})^2 + (\gamma_{\om {\bf
q}_{\perp}}(k))^2}\label{r-equil}  
\end{eqnarray}

Comparing Eq.~(\ref{mp-zero}) with (\ref{mp-final}),
(\ref{r-equil}) we see, 
that the effect of collisions, which makes the largest contribution to
$K$ can be described as $\delta(\om - \epsilon_{{\bf q}_{\perp}k})
\Rightarrow \delta_{\gamma}(\om - \epsilon_{{\bf q}_{\perp}k})$, as can
be expected on the intuitive grounds. Now we substitute
Eqs.~(\ref{mp-final}), (\ref{r-equil}) into  
(\ref{rate-zero}) and integrate $\int d\om/2 \pi$ by using
the result of~\cite{tails-galitski} for large momenta, $q \gg q_T$:
\begin{eqnarray}
\int \frac{d \om}{2 \pi} n_M(\om) \delta_{\gamma}(\om - \epsilon_{{\bf
q}_{\perp} k})&=&n_M(\epsilon_{{\bf q}_{\perp} k}) + \delta
n_{\gamma}({\bf q}_{\perp}k), \nonumber
\end{eqnarray}
where $\delta n_{\gamma}({\bf q}_{\perp}k)$ is the ``quantum tail''. 

To describe the tunneling of particles $m_1$ and $m_2$
we  perform the same steps as those leading
to Eq.~(\ref{rate-zero-final}). Note that this procedure implies
quantization of the hypothetical particle with the mass equal to the
reduced mass of fusing particles. Strictly speaking, this
approximation is correct in the limit when one particle is light
and the other is heavy. Keeping this in mind, we arrive at the
answer for $K$: 
\begin{eqnarray*}
K=n_0 \int \frac{d \om_{\bf p}}{2 \pi} \int_0^{\infty} \frac{4 \pi p^2 dp}{(2 \pi)^3} \frac{S(\bar{\epsilon}_p)}{\bar{\epsilon}_p}&& \hspace*{25mm}
\end{eqnarray*}
\begin{equation}
\frac{p}{m_r} W(p) \bar{n}_M(\om) \delta_{\gamma}(\om - \epsilon_{\bf p})
\label{k-delta} 
\end{equation}
We see that it involves integrals over both the energy and
momentum. The width of the resonance, $\gamma(\om_{\bf p}, {\bf p})$, 
is, in general, a complicated function of energy and momentum. It 
obeys under certain assumptions the integral
equation~\cite{tails-cross-section-star} which follows from the Dyson diagram
expansion. More simplified calculations,
based on the use of the zero-th order Green functions, lead to the analytical
answer found in ~\cite{tails-galitski} (see Appendix A).
We use this answer for $\gamma$~\cite{tails-galitski,tails-cross-section-star}
and obtain
\begin{eqnarray*}
K=n_0 \int_0^{\infty} \frac{4 \pi p^2 dp}{(2 \pi)^3} 
\frac{S(\bar{\epsilon}_p)}{\bar{\epsilon}_p}&& \hspace*{25mm}
\end{eqnarray*}  
\begin{equation}
\left[\bar{n}_M(p) +
\delta \bar{n}_{\gamma} (p) \right]\equiv K_M + K_{\gamma}. \label{k-semifinal}
\end{equation}
Finally, we evaluate $\Sigma^R$
through the collision frequency, $\nu_T$, 
carry out the integral in
(\ref{k-semifinal}) and form the ratio $r_{12}=K_{\gamma}/K_M$ (see
Appendices A and B):
\begin{equation}
r_{12}=\frac{3^{19/2}}{8 \pi^{3/2}} \sum_{j} \frac{h
\nu_T(m_{coll})}{T} \left(\frac{m_{coll}}{m_r}\right)^{2}
\frac{e^{\tau_{12}}}{\tau_{12}^8} \label{ratio-final}
\end{equation}
Here $m_{coll}=m_r m_j/(m_r + m_j)$, $m_r=m_1 m_2/(m_1 + m_2)$,
$\tau_{12}=3(\pi/2)^{2/3} 
\left(\frac{100 Z_1^2 Z_2^2 A_{12}}{T_k}\right)^{1/3}$, $A_{12}=A_1
A_2/(A_1 + A_2)$, $T_k$ is the temperature in keV; $A_1$, $A_2$ are
the atomic numbers of tunneling particles, 
$m_j$ is the mass of the
background particles colliding with tunneling particles. Summation is
over all species of colliding background particles, including
tunneling species.
This result, (\ref{k-semifinal}), (\ref{ratio-final}), supports the idea
of averaging of $\sigma v$ over momentum 
distribution postulated in ~\cite{neutrino-vlad}.

Note an important feature of the result (\ref{ratio-final}): this
ratio goes to zero as $\hbar \nu_{\bf p}/T \rightarrow 0$. As we show
in Appendix A, the width, $\gamma$, can be approximated as $\hbar
\nu_{\bf p}$~\cite{tails-cross-section-star}. Therefore, this means
that in the limit of negligible 
width, the correction to the rate becomes very small, and the
reaction rate coincides with the usual Gamow rate. This is expected,
since the smaller the width, the closer the particle to the mass
shell. And the Gamow rate is derived for on-shell particles. 
On the other hand, the ratio becomes large, if $\hbar \nu_{\bf p}/T$ is not
very small, yet still $\hbar \nu_{\bf p}/T<1$. It is this condition,
which allows us to use the result of \cite{tails-galitski} for
$\gamma$ in Eqs.~(\ref{k-semifinal}), (\ref{ratio-final}). In the
opposite case, 
$\hbar \nu_{\bf p}/T\geq1$, corresponding to more strongly coupled plasma, the
theory presented here does not apply.

Consider a $DT$ plasma of density
$\rho=10\, {\rm g/cm}^3$ and temperature $T=0.1$~keV. The ratio $\hbar
\nu_{\bf p}/T=0.0075$ for these conditions. 
Then
Eq.~(\ref{ratio-final}) will lead to the rate accelerated by five orders
of magnitude as compared to the conventional answer. We find the ratio
$r_{DT}=K^{\gamma}_{DT}/K^M_{DT}=1.8 \cdot 10^5$, with
$K^M_{DT}=2.6 \cdot 10^{-30} {\rm cm}^3/{\rm sec}$,
$K^{\gamma}_{DT}=4.8 \cdot 10^{-25} {\rm cm}^3/{\rm sec}$.

It is also interesting to consider how the reaction rates are changed 
under conditions relevant to astrophysics, e.g. in the Sun
interior. Modification of several reaction rates and their influence on
neutrino fluxes from the Sun was considered in
~\cite{neutrino-vlad}. Here we give examples of $p+Be^7$ and $p+p$
reactions, occuring at
the center of the Sun, $r=0$, where the temperature is $T=
1.336$~keV and the density $\rho=156~{\rm g}/{\rm cm}^2$, according to
the solar 
model of~\cite{solar-bahcall-767}. The estimate for $\gamma_{17}/T$ is
$\hbar \nu_T/T\sim 0.0056$, while the rates and the ratio, $r_{17}$, are
$K^M_{17}=9.6\cdot 
10^{-36}{\rm cm}^3/{\rm sec}$, $K^{\gamma}_{17}=4.0\cdot 10^{-30} {\rm
cm}^3/{\rm sec}$, $r_{17}=4.17\cdot 10^{5}$.

In case of heavier elements and not very small $\hbar \nu_T/T$, the
ratio $r_{ij}$ 
becomes much higher. On the other hand, the corresponding numbers for
the $p+p$ reaction under the same conditions are: $\gamma/T\sim \hbar
\nu_T/T\sim 0.00054$, $K^M_{11}=1.4\cdot 10^{-43} {\rm cm}^3/{\rm
sec}$, $r_{11}=3\cdot 10^{-3}$. The reaction rate is essentially the
same. Note that $\hbar \nu_T/T\sim 0.00054$ is rather small. This confirms
the conclusion of the discussion above, that small $\gamma$ lead to
insignificant changes in the reaction rates. Note, however, that we
need to have some gauge to determine which $\gamma$ is small. Naturally,
such a gauge should be a ratio of the height of the Coulomb barrier to the
average particle energy. As we see from Eq.~(\ref{ratio-final}) this
ratio enters the answer for $r_{12}$ through $\tau_{12}$.

In conclusion, we derived the momentum distribution in equilibrium through
real-time technique of Keldysh and Korenman which coincides with the
distribution, obtained in imaginary time~\cite{tails-galitski}. This
distribution takes into account off-shell effects which lead to
the quantum tail with power-like dependence on momentum
in contrast to the distribution for on-shell particles which is
Fermi-Dirac or Maxwellian. We calculated the rates of fusion
reactions for particles in a weakly coupled plasma and
showed that off-shell
effects may significantly enhance the reaction rates.

I would like to acknowledge helpful discussions with N.~J.~Fisch,
D.~Spergel, A.~N.~Starostin, A.~Demura, A.~A.~Panteleev.
This work was supported by NSF--DOE grant
DE-FG02-97ER54436.

E-mail: {\it vsavchen\@pppl.gov}

\end{multicols}

\vspace*{3mm}

\begin{centering}

{\bf APPENDIX A

CALCULATION OF THE MASS OPERATOR, $\Sigma^R$,
AND THE DISTRIBUTION FUNCTION OVER MOMENTUM}

\end{centering}

\vspace*{3mm}

\begin{multicols}{2}
\dimen100=\columnwidth \setlength{\columnwidth}{3.375in}

Galitski and Yakimets found how off-shell processes change the
distribution function over momentum in equilibrium, using imaginary
time technique~\cite{tails-galitski}.
In this Appendix we show how to find the same result through real time
diagram method. We also generalize it slightly by allowing for
collisions between different types of particles.

Dyson equations for the function $G^{-+}$ can be written as~\cite{landau-10}
\begin{eqnarray}
&&\left(i \frac{\partial}{\partial t} + i \frac{\bf p}{m}\cdot
\frac{\partial}{\partial {\bf R}}\right) G^{-+}\left({\bf R}, t;\om,
{\bf p}\right)= \nonumber\\
&&-\Sigma^{-+} G^{+-} + \Sigma^{+-} G^{-+}
\label{dyson-general}
\end{eqnarray}
To simplify notation we do not write explicitely the variables the
functions $\Sigma$ and $G$ depend upon.
We set to zero the LHS of this equation in equilibrium and obtain
\begin{equation}
G^{-+}=\frac{\Sigma^{-+}}{\Sigma^{+-}} \left(G^R - G^A + G^{-+}\right),
\label{dyson-equilib}
\end{equation}
which follows from the relation between Green
functions~\cite{landau-10}. 
The result for $G^{-+}$ follows by substituting
Eq.~(\ref{dyson-equilib}) into (\ref{gr}):
\begin{equation}
G^{-+}=- 2 \pi i n_F(\om) \frac{Im \Sigma^R}{(\om - \epsilon_{\bf p} - Re
\Sigma^R)^2 + (Im \Sigma^R)^2}
\label{gmp-equilibrium}
\end{equation}
We use this equation to find the distribution function over momentum
which is
\begin{equation}
f({\bf p})=\int\limits_{-\infty}^{\infty} \frac{d\, \om_{\bf p}}{2\pi}
G^{-+}(\om_{\bf p}, {\bf p}).
\label{fp-def}
\end{equation}
First we calculate the
mass-operator $\Sigma^R$. It can be expressed 
using the relation 
\begin{equation}
\Sigma^R=\frac{1}{2}(\Sigma^{--} - \Sigma^{++}) -
\frac{1}{2}(\Sigma^{+-} - \Sigma^{-+})
\label{sigmar}  
\end{equation}
The first term in this equation corresponds to the real part $Re
\Sigma^R$, while the second term is imaginary part $Im
\Sigma^R$. Since it is the imaginary part of $\Sigma^R$ which is
responsible for the quantum tail~\cite{tails-galitski} we calculate
only $Im \Sigma^R$ and ignore $Re \Sigma^R$ altogether.

\begin{figure}[!h]
         \centering
         \mbox{\epsfig{figure=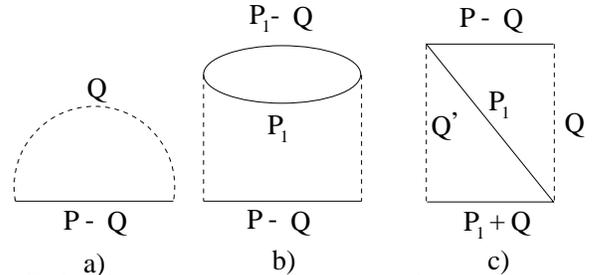, width=3.0
         true in}} 
         \caption{First and second-order terms in the diagram
         expansion of $\Sigma^R$.} 
         \label{sigma-terms} 
\end{figure}

The first two
terms of the series for the operators $\Sigma^{\alpha \beta}$ are
shown in the Fig.~\ref{sigma-terms}. One has to assign all
possible combination of signs $-,+$ to the vertices, keeping in mind
that the sign does not change along the dashed
line~\cite{statistical-abrikosov}. 

In a more rigorous approach,
based on the resummation of the series of graphs,
one can
show~\cite{trydi-galitski,statistical-abrikosov,statistical-mechanics-kadanov}
that, solid
lines in the diagrams of Fig.~\ref{sigma-terms} correspond to {\em exact} 
Green-functions, while the dashed lines represent scattering
amplitudes of the particles in the media. However, such resummations lead
to intractable non-linear integro-differential equations.

As a first approximation we assume that, the
solid lines correspond to 
zero-order Green functions, while the dashed lines correspond to the
interaction potential between the particles.

These  diagrams represent quite different physical
phenomena. Diagram ``a)'' describes the screening effects and
formation of the quasiparticles in the plasma. Therefore, an imaginary
part of this 
graph represents decay of the quasi-particles. 
This term is
conventionally put into the LHS of the kinetic equation, signifying
that it has nothing to do with the collisional integral~\cite{statistical-mechanics-kadanov}. Diagram ``b)'', on the
other hand, if evaluated with zero-order Green-functions, gives the
usual Boltzman collision
terms~\cite{landau-10,statistical-mechanics-kadanov}. 
Therefore, the imaginary parts of the diagrams ``b)'' and ``c)''
describe {\it damping} of the oscillators used for the second
quantization of the {\it interacting
particles}.

We evaluate the ``b)'' and ``c)''-type of diagrams for $\Sigma^{-+}$,
assuming that collisions occur between particles of type $a$ and
$b$. Therefore, we assign the index $a$ to the Green function,
represented by the horizontal line in the diagram ``b)'', and the
index $b$ to the Green functions of the loop. Note that since the
diagram ``c)'' corresponds to the exchange processes only single index
can be assigned to its Green functions. This obviously reflects the
fact that only identical particles are subject to exchange.
Taking into account these remarks, we write the analytical expression
for the diagramm b) in the form 
\begin{eqnarray}
-i\Sigma^{-+}=\int G^{-+}_a&&(P-Q) G^{+-}_b(P_1) G^{-+}_b(P_1 - Q) \nonumber\\
&&\left|V({\bf q})\right|^2 \frac{d\om_{{\bf p}_1} d{\bf p}_1}{(2
\pi)^4} \frac{d\om_{{\bf q}} d {\bf q}}{(2
\pi)^4}, \label{sigma-mp}
\end{eqnarray}
where capital letters denote four-momentum, e.g. $P=(\om_{\bf p}, {\bf
p})$ and
$V({\bf q})$ is the interaction potential between the particles.
We substitute here equilibrium zero-th order Green functions~\cite{landau-10}
\begin{equation}
G^{-+}(\om_p {\bf p})=2 \, \pi \, i \; n_{\bf p} \;
\delta(\om_p - \epsilon_{\bf p} + \mu), \label{gmp-delta}
\end{equation}
\begin{equation}
G^{+-}(\om_p {\bf p})=-2 \, \pi \, i \; (1 - n_{\bf p})
\delta(\om - \epsilon_{\bf p} + \mu). \label{gpm-delta}
\end{equation}
Analogously, we obtain an expression for the diagram c). 

For a moment, we omit the indices $a$ and $b$ to simplify the formulas,
but write them explicitely in the final formulas.
Then the result for $\Sigma^{-+}$ is 
\begin{eqnarray}
&&\Sigma^{-+}(\om_{\bf p}, {\bf p})=(-2\pi i) \int\frac{d\,{\bf
p}_1}{(2 \pi)^3}  
\int\frac{d\,{\bf q}}{(2 \pi)^3} \nonumber\\
&&\left(2V_{\bf q}^2 - V_{\bf q}
V_{{\bf q}-{\bf p}-{\bf p}_1}\right) 
n_{{\bf p}-{\bf q}} n_{{\bf
p}_1-{\bf q}} \nonumber\\
&&\delta(\om_{\bf p} - \epsilon_{{\bf p} - {\bf q}} +
\epsilon_{{\bf 
p}_1} - \epsilon_{{\bf p}_1 - {\bf q}}) \label{sigmamp-wexchange}
\end{eqnarray}

Proceeding in a similar fashion with $\Sigma^{+-}$ and using
Eq.~(\ref{sigmar}) we obtain the final
result for $Im \Sigma^R\equiv \gamma(\om_p, {\bf p})$:
\begin{eqnarray} 
\gamma&&(\om_{\bf p}, {\bf p})=\pi\, i \int \frac{d\,{\bf q}}{(2\pi)^3}
\, \frac{d\,{\bf p}_1}{(2\pi)^3} V_{\bf q} (2 V_{\bf
q} - V_{{\bf q} - {\bf p} - {\bf p}_1}) \nonumber\\
&&\left[n_{{\bf p} - {\bf q}}
(n_{{\bf p}_1 - {\bf q}} - n_{{\bf p}_1}) + \right.
\left.n_{{\bf p}_1}(1 - n_{{\bf p}_1 - {\bf
q}})\right] \nonumber\\
&&\delta(\om_{\bf p} - \epsilon_{{\bf p} - {\bf q}} + \epsilon_{{\bf
p}_1} - \epsilon_{{\bf p}_1 - {\bf q}}) \label{gamma-galitski}
\end{eqnarray}
This answer coincides with the result of
\cite{sigma-galitski,tails-galitski} obtained in the imaginary-time
technique. 

Note that we could improve this relation by substituting
$\delta_{\gamma}$ of Eq.~(\ref{r-equil}) instead of the
delta-function. This, however, would lead to
an integral equation for $\gamma$, and its first iteration will be
just Eq.~(\ref{gamma-galitski}).

Now we can find the distribution function over momentum from
Eqs.~(\ref{fp-def}), (\ref{gamma-galitski}), (\ref{gmp-equilibrium}) 
\begin{equation}
f({\bf p})=\int\limits_{-\infty}^{\infty} \frac{d\,
\om_{\bf p}}{(2\pi)^2} \frac{n_F(\om) \gamma(\om_{\bf
p}, {\bf p})}{(\om_{\bf p} - \epsilon_{\bf p})^2 + \gamma^2(\om_{\bf p}, {\bf p})}  \label{fp-definition}
\end{equation}
by substituting Eq.~(\ref{gamma-galitski}) into
Eq.~(\ref{gmp-equilibrium})~\footnote{From now on we follow closely
the work of \cite{tails-galitski}}. The integral $\int d\om$ can be rewritten
as a sum of two terms
\begin{eqnarray}
f({\bf p})=&&\int\limits_{-\infty}^{\epsilon_0} \frac{d\, \om_{\bf
p}}{2\pi} n_F(\om) \delta_{\gamma}(\om_{\bf p} - \epsilon_{\bf p}) +
\nonumber\\ 
&&\int\limits_{\epsilon_0}^{\infty} \frac{d\, \om_{\bf 
p}}{2\pi} n_F(\om) \delta_{\gamma}(\om_{\bf p} - \epsilon_{\bf p}),
\label{two-terms} 
\end{eqnarray}
where $\epsilon_0$ is the characteristic energy. In the degenerate
case, $T\ll \mu$, we have $\epsilon_0=\mu$, in the nondegenerate case,
$T\gg \mu$ and $\epsilon_0=T$. We define $p_0$ as $\epsilon_0=p_0^2/2m$.

Consider the case of large momenta, $p\gg p_0$. The function
$\delta_{\gamma}(\om_{\bf p} - \epsilon_{\bf p})$ has a maximum at
$\om_*=\epsilon_{\bf p}\gg \epsilon_0$, so that $\om_*
\in[\epsilon_0,\infty]$, and the contribution of the second integral
can be approximated as $n_{\bf p}$. In the region of integration of
the first integral in Eq.~(\ref{two-terms}) we can set
$n_F(\om)\approx 1$. Then 
Eq.~(\ref{two-terms}) becomes for $p\gg p_0$
\begin{equation}
f({\bf p})=n_{\bf p} + \int\limits_{-\infty}^{\epsilon_0} \frac{d\, \om_{\bf
p}}{2\pi} \frac{\gamma(\om_{\bf
p}, {\bf p})}{(\om_{\bf p} - \epsilon_{\bf p})^2}\equiv n_{\bf p} + \delta
n_{\gamma}({\bf p}), 
\label{fp-semifinal} 
\end{equation}
We carry out the integral in the second term of (\ref{fp-semifinal}),
use Eq.~(\ref{gamma-galitski}) and
obtain 
\begin{eqnarray}
&&\delta n_{\gamma}({\bf p})=\int \frac{d\,{\bf q}}{(2\pi)^3}
\, \frac{d\,{\bf p}_1}{(2\pi)^3} V_{\bf q} (2 V_{\bf
q} - V_{{\bf q} - {\bf p} - {\bf p}_1}) \nonumber\\
&& \frac{\left[n_{{\bf p} - {\bf q}}
(n_{{\bf p}_1 - {\bf q}} - n_{{\bf p}_1}) + 
n_{{\bf p}_1}(1 - n_{{\bf p}_1 - {\bf
q}})\right]}{(\epsilon_{\bf p} - \epsilon_{{\bf p} - {\bf q}} + \epsilon_{{\bf
p}_1} - \epsilon_{{\bf p}_1 - {\bf q}})^2}, \label{deltaf}
\end{eqnarray}
with the condition on momenta 
\begin{equation}
-\infty<\epsilon_{{\bf p} - {\bf q}} - \epsilon_{{\bf
p}_1} + \epsilon_{{\bf p}_1 - {\bf q}}<\epsilon_0 \label{om-condition}
\end{equation}
due to the integration of the delta-function from
$\gamma(\om_{\bf p}, {\bf p})$.

The integral in Eq.~(\ref{deltaf}) should be carried out
numerically. We can, however, obtain a reasonable estimate in the case
of large momentum, $p\gg p_0$. Observe that the first two terms in the
brackets $[..]$ should have the same dependence on $p$.
To obtain an estimate consider the product of the
first term in $(..)$, $V_{\bf q}^2$, and the first term in
$[..]$ which involves the product $n_{{\bf p} - {\bf q}} n_{{\bf p}_1
- {\bf q}}$. We denote it $I_1({\bf p})$.
Note that there is a region of non-exponential
contribution to the integral, which is limited to the region of momenta
\begin{equation}
\left|{\bf p} - {\bf q}\right|\sim \left|{\bf p}_1 - {\bf
q}\right|\sim p_0, p\sim p_1 \sim q \gg p_0
\label{p-equality} 
\end{equation}
The inequality
(\ref{om-condition}) is satisfied which, in turn, means that the
dominant contribution to this product can be approximated as 
\begin{equation}
I_1({\bf p})=\int \frac{d\,{\bf q}}{(2\pi)^3}
\, \int \frac{d\,{\bf p}_1}{(2\pi)^3} 2 V_{\bf q}^2 
\frac{n_{{\bf p} - {\bf q}}^{(a)} n_{{\bf p}_1 - {\bf
q}}^{(b)}}{\tilde{\epsilon}_{\bf p}^2},  
\end{equation}
where $\tilde{\epsilon}_{\bf p}\equiv {\bf p}^2/2 m_{ab}$ and we
reinstalled indices $a$ and $b$.
Consider non-degenerate case. 

We change variables, ${\bf p} - {\bf q}={\bf q}'$, ${\bf p}_1 - {\bf
q}={\bf p}_1'$ and obtain an estimate
\begin{equation}
I_1({\bf p})=e_a^2 e_b^2 \frac{(2 m_{ab})^2}{p^8}
\left(\frac{m_a}{2\beta \pi}\right)^{3/2} \left(\frac{m_b}{2\beta
\pi}\right)^{3/2} e^{\beta (\mu_a + \mu_b)}
\label{ip-intermed}
\end{equation}
We substitute the relation
\begin{equation}
e^{\beta (\mu_a + \mu_b)}=n^{(a)} \lambda_a^3 n^{(b)} \lambda_b^3
\end{equation}
\begin{equation}
\lambda_a^{-3}=\left(\frac{m_a}{2\beta \pi}\right)^{3/2}, \;\; 
\lambda_b^{-3}=\left(\frac{m_b}{2\beta \pi}\right)^{3/2}
\end{equation}
into Eq.~(\ref{ip-intermed}) and find
\begin{equation}
I_1({\bf p})=\frac{(2 m_{ab})^2}{p^8}  (e_a e_b)^2 n^{(a)} n^{(b)}
\label{ip-inter}
\end{equation}
Since the physical reason for non-zero $\gamma(\om_p, {\bf p})$ is
collisions, we rewrite (\ref{ip-inter}) in terms of the collision
frequency
\begin{equation}
\nu_p=\left(\frac{\pi}{2}\right)^{3/2} n^{(a)}(e_a e_b)^2
m_{ab}^{-1/2} T^{-3/2} \frac{(2m_{ab} T)^{3/2}}{p^3}
\end{equation}
as 
\begin{equation}
I_1({\bf p})=\frac{2}{\pi^{3/2}}\frac{\hbar \nu_p
T}{\tilde{\epsilon}_{\bf p}^2} \frac{\tilde{p}_T}{p}
\frac{n^{(b)}}{\tilde{p}_T^3},
\label{ip}
\end{equation}
where $\tilde{p}_T\equiv (2m_{ab}T)^{1/2}$.
This is essentially the result of~\cite{tails-galitski} for the
degenerate case, sligtly generalized
to the case of colliding particles with different masses. 

In the non-degenerate case we can keep only the term linear in $n_{\bf
p}$ in Eq.~(\ref{deltaf}), because the quadratic terms are smaller. 
We denote this term by $I_3$.
Since there is no exponential suppression in the $q$-integral for this
term, we have to evaluate the limits of the $q$-integration
carefully. We can obtain these limits from
Eq.~(\ref{om-condition}). After some algebra it leads to
\begin{equation}
\frac{q^2}{2m_{ab}} - q x\left|\frac{\bf p}{m_a} + 
\frac{{\bf p}_1}{m_b}\right| > \epsilon_{\bf p} - T,
\label{q-condition}
\end{equation}
where $x$ is the cosine of the angle between ${\bf q}$ and $\frac{\bf
p}{m_a} + \frac{{\bf p}_1}{m_b}$.
We resolve this inequality in the limit of large $p\gg p_T$,
neglecting all terms on the order of $p_T^2$ in comparison to
$p^2$. Note that since the term in quesiton, $I_3$, is proportional to
$n_{{\bf p}_1}$, the contribution from the region $p_1\geq p_T$ is
exponentially suppressed. Since we are interested in the
non-exponential contribution only, we can assume that $p_1\sim p_T$ in
Eq.~(\ref{q-condition}). Then we finally arrive at
\begin{equation}
q> q_*=\frac{m_{ab}}{m_a} p(1+\sqrt{1+\frac{m_{a}}{m_{ab}}})
\label{q-cond}
\end{equation}

To obtain an estimate for $I_3$, we carry out the integral of the
linear term in $n_{{\bf p}_1}$ from Eq.~(\ref{deltaf}), multiplied by
the first term in $(..)$, $V_{\bf q}^2$:
\begin{eqnarray}
&&I_3({\bf p})=2 \pi \int d {\bf p}_1 \int\limits_{-1}^{1} dx
\int\limits_{q_*}^{\infty} q^2 dq \nonumber\\
&&\frac{(e_a e_b)^2}{q^4}
\frac{e^{-p_1^2/2m_b T} e^{\beta
(\mu_a + \mu_b)}}{\left(\frac{q^2}{2m_{ab}} - q x \left|\frac{\bf p}{m_a} + 
\frac{{\bf p}_1}{m_b}\right|\right)^2}
\label{i3-x} 
\end{eqnarray}
After performing the integral over $x$, we are left with the integral
over $y$, which is the cosine of the angle between ${\bf p}$ and ${\bf
p}_1$. The $y$-dependent term becomes 
\begin{equation}
\int dy \frac{1}{\left(\frac{q^2}{2m_{ab}}\right)^2 -
q^2\left(\frac{p^2}{m_a^2} + \frac{p_1^2}{m_b^2} + \frac{2 p p_1
y}{m_a m_b}\right)}
\label{int-y}
\end{equation}
We carry out this integral, neglect all the terms of the order of
$p_T^2$ in comparison to $p^2$ and obtain $2m_a^2/(q^2 p^2)$. The
remaining integrals over $\int\limits_0^{\infty} d p_1$ and
$\int\limits_{q_*}^{\infty} d q$ are trivial. Using Eq.~(\ref{q-cond})
we get the result for $I_3$ 
\begin{eqnarray*}
&&I_3(p)=\frac{2}{3} \frac{1}{(2\pi)^3} \frac{m_a^2}{p^5}
\left(\frac{m_a}{m_{ab}}\right)^3
\frac{(e_a e_b)^2 n^{(b)} e^{\beta
\mu_a}}{\left(1+\sqrt{1+\frac{m_a}{m_{ab}}}\right)^3}  
\label{i3-final}
\end{eqnarray*}

Now we can see that different terms in Eq.~(\ref{deltaf}) have a
different dependence on $p$. Note however that the answers for these terms are
really estimates. To find the correct answer one should solve non-linear
equation for $\gamma$~\cite{tails-cross-section-star}, which is
possible only through numerical methods. 

It is also possible to obtain an estimate through dimensional
analysis which leads to~\cite{tails-cross-section-star} 
\begin{equation}
\gamma=2 Im \Sigma^R \sim 2 \hbar \nu_{\bf p}
\label{gamma-estimate}
\end{equation}
This expression has to be substituted into
Eq.~(\ref{fp-semifinal}). Note that we have to carry out the integral
over energy from $-\infty$ to $T$. Since we do
not know how $\gamma$ behaves at negative energies, we can only make an
assumption that, this region does not contribute significantly to the
integral, and the main contribution comes from $\om_{\bf p} \sim
T$. Then the integral is equal roughly to 
\begin{equation}
I(p)=\frac{\hbar \nu_p T}{2\pi \epsilon_{\bf p}^2} \frac{n^{(a)}}{p_T^3}.
\label{ip-star}
\end{equation}
We can simplify it to 
\begin{equation}
I(p)=\frac{\hbar \nu_p T}{2\pi \epsilon_{\bf p}^2} e^{\beta \mu_a}
\label{ip-star-final}
\end{equation}
which is the result for the non-degenerate case found in
~\cite{tails-cross-section-star}. We will use this estimate in our
calculations of the reaction rate because of its simplicity.

\end{multicols}

\vspace*{3mm}

\begin{centering}

{\bf APPENDIX B

REACTION RATE: CALCULATION OF THE INTEGRAL OVER $\delta
n_{\gamma}({\bf p})$}

\end{centering}

\vspace*{3mm}

\begin{multicols}{2}
\dimen100=\columnwidth \setlength{\columnwidth}{3.375in}

In this Appendix we find the rate $K_{\gamma}$ of Eq.~(\ref{k-semifinal}). 
Observe, that in the main text we used approximation of a hypothetical
particle with the mass equal to the reduced mass of fusing particles,
$m_r$. It is this particle which undergoes collisions with the
background, leading to its power-like distribution over
momentum. 
We will need the result (\ref{ip-star}) for these particles. We write as
\begin{equation}
I({\bf p})=\frac{\hbar \nu_p
T}{2 \pi \tilde{\epsilon}_{\bf p}^2} 
\left(\frac{m_r}{m_{coll}}\right)^{3/2}
e^{\beta \mu_a}
\label{ip-final}
\end{equation}
taking into account that
\begin{equation}
\frac{n^{(a)}}{\tilde{p}_T^3}=\left(\frac{m_r}{m_{coll}}\right)^{3/2}
e^{\beta \mu_a}.
\end{equation}

Therefore, we substitute Eq.~(\ref{ip-final}) into
Eq.~(\ref{k-semifinal}) 
and obtain $K_{\gamma}$ 
\begin{eqnarray} 
&&K_{\gamma}= \left(\frac{m_r}{m_{coll}}\right)^{3/2}
(2m_r)^2 
\frac{\hbar \nu_T(m_{coll}) T}{2 \pi} \nonumber\\
&&S(\epsilon^*)(2m_{coll} T)^{3/2}
\int\limits_0^{\infty} e^{-\pi p_G/p} \frac{1}{p^6} dp,
\label{k-int}
\end{eqnarray}
where $S(\epsilon^*)$ is the astrophysical factor at some fixed value of energy.
The integral is trivial and is equal to
\begin{equation} 
\int\limits_0^{\infty} e^{-\pi p_G/p} \frac{1}{p^6}
dp=\frac{24}{\pi^5 p_G^5}.
\label{pg-integral}
\end{equation}
For the sake of completeness, we also find here the usual Gamow rate:
\begin{equation} 
K_M=4\pi S(\epsilon^*)\int_0^{\infty} e^{-\pi p_G/p - p^2/2m_r T} p dp.
\end{equation}
The result, obtained by using the steepest descend method, is
\begin{equation} 
K_M=S(\epsilon^*) \frac{2}{3} \pi^{1/2} (m_r T) \tau^{1/2} e^{-\tau}.
\label{k-m}
\end{equation}

Strictly speaking, the astrophysical factors in Eqs.~(\ref{k-int}) and
(\ref{k-m}) need not be the same. In fact, one should consider a
generalized astrophysical factor, $S(\om_p, \epsilon_{\bf p})$,
depending both on energy and momenta. Such factors would take into
account the influence of the
off-shell effects on nuclear transformations. However, as
far as we know, they have not been studied. Therefore, we assume that
the factors in Eqs.~(\ref{k-int}), (\ref{k-m}) are the
same. 

With this in mind, we can form the ratio
\begin{equation} 
r_{12}=\frac{K_{\gamma}}{K_M}=\frac{3^{19/2}}{2^2 \pi^{3/2}}
\frac{\hbar \nu_T(m_{coll})}{T} \left(\frac{m_{coll}}{m_r}\right)^{2}
\frac{e^{\tau}}{\tau^{8}}
\label{ratio-eight}
\end{equation}


\end{multicols}

\end{document}